\documentclass[useAMS,usenatbib,usegraphicx]{mn2e}

\usepackage{times}
\usepackage{multicol}

\def\gs{\mathrel{\raise0.35ex\hbox{$\scriptstyle >$}\kern-0.6em
\lower0.40ex\hbox{{$\scriptstyle \sim$}}}}
\def\ls{\mathrel{\raise0.35ex\hbox{$\scriptstyle <$}\kern-0.6em
\lower0.40ex\hbox{{$\scriptstyle \sim$}}}}

\defcitealias{richards00}{R00}

\title{A catalogue of $\mu$Jy radio sources in northern legacy fields}
\author[Biggs \& Ivison]{A.~D.~Biggs$^{1}$\thanks{E-mail: adb@roe.ac.uk} and R.~J.~Ivison$^{1,2}$\\
$^1$UK Astronomy Technology Centre, Royal Observatory, Blackford Hill, Edinburgh EH9 3HJ\\
$^2$Institute for Astronomy, University of Edinburgh, Blackford Hill, Edinburgh EH9 3HJ}

\begin{document}

\maketitle

\begin{abstract} We present catalogues of faint 1.4-GHz radio sources from
extremely-deep Very Large Array pointings in the Lockman Hole, the Hubble Deep
Field North (HDF-N) and ELAIS~N2. Our analysis of the HDF-N data has produced maps
that are significantly deeper than those previously published and we have used these
to search for counterparts to submm sources.
For each of the fields we have derived normalised differential source counts and in the
case of the HDF-N find no evidence for the previously-reported under-density of sources;
our counts are entirely consistent with those found for the majority of
other fields. The catalogues are available as an online supplement to this
paper and the maps are also available for download.
\end{abstract}


\section{Introduction}

Our knowledge of the evolution of galaxies and active galactic nuclei (AGN)
arguably owes as much to radio surveys as to those in any other band. Our
knowledge of the cosmological evolution of radio-loud AGN is an obvious example
\citep{dunlop90}, and one which foretold with remarkable accuracy the
history of accretion luminosity, and even that of star formation \citep{dunlop98}.

Recent work has shown that deep ($\mu$Jy) radio surveys have the power to probe
star-forming galaxies, unbiased by dust obscuration, out to very high
redshifts. Moreover, deep radio data have proved to be extraordinarily
complementary to datasets obtained in the submillimetre (submm), infrared (IR)
and X-ray wavebands \citep{ivison02,donley05}.

Unfortunately, the radio waveband has not benefited from the technological
improvements and investment seen at shorter wavelengths. This is in
stark contrast with our mapping capability in the submm and IR wavebands, which
has improved by many orders of magnitude in barely a decade.

At 20\,cm, the National Radio Astronomy Observatory's (NRAO's) Very Large Array
(VLA) has been the dominant survey facility for several decades, with the
Australia Telescope (ATCA), MERLIN and Westerbork (WSRT) synthesis arrays all
having found significant niches. The most recent major radio surveys were
the NRAO VLA Sky Survey (NVSS), covering 82~per~cent of the sky to
$\sigma = 450\,\mu$Jy\,beam$^{-1}$ \citep{condon98} and FIRST (Faint Images of the
Radio Sky at Twenty\,cm, the radio equivalent of the Palomar
Observatory Sky Survey) which mapped 10$^4$\,deg$^2$ to around
$\sigma = 140\,\mu$Jy\,beam$^{-1}$ \citep{white97}.

Going deeper, radio surveys in a plethora of fields have reached noise levels
of $\sigma \sim$10--20\,$\mu$Jy\,beam$^{-1}$, the confusion limit at 20\,cm for
arrays smaller than $\sim$5\,km. These surveys, often unpublished but totalling
around 10\,deg$^2$, have been carried out with a variety of instruments (predominantly
the VLA, but also the WSRT and ATCA) and include the Hubble Deep Field North
\citep{richards00}, the Hubble Deep Field South \citep{huynh05}, the
{\em Spitzer} First Look Survey \citep{condon03,morganti04}, the VLA--VIRMOS
Deep Field \citep{bondi03} and the Phoenix Deep Survey \citep{hopkins03}.

Given the rarity with which surveys have probed significantly below the
$\sigma \sim 10\,\mu$Jy\,beam$^{-1}$ level \citep{ivison02,muxlow05} it is
natural to assume that to do so must require enormous allocations
of observing time. This is not the case. Here, we describe three radio
pointings which, together, required less than one week of VLA time. Each covers
an area of $\sim$320\,arcmin$^2$ with 1.5-arcsec resolution, two of them to
a noise level of $\sim$5\,$\mu$Jy\,beam$^{-1}$, the other to
$\sim$10\,$\mu$Jy\,beam$^{-1}$.

Two of the VLA pointings lie within two regions of the sky which have very low
Galactic backgrounds ($<$0.5\,MJy\,sr$^{-1}$ at 100\,$\mu$m and $E(B-V)< 0.01$) and
which were selected for the {\em Spitzer} Wide-area InfraRed Extragalactic (SWIRE)
survey. These are the `Lockman Hole'
and ELAIS N2. The Lockman Hole is a $4\times3$-deg$^2$ region centred at
$\mathrm{10^h 48^m, +57^{\circ} 04'}$ (J2000) with
$\sim5 \times 10^{19}$\,cm$^{-2}$ of H\,{\sc i} \citep*{lockman86}. This makes it
uniquely well suited to panchromatic, deep studies of the
Universe. The final VLA pointing is located in the Hubble Deep Field North
(HDF-N), part of the Great Observatories Origins Deep Surveys \citep[GOODS,][]{dickinson03}.
The ELAIS N2 and HDF-N data have been previously discussed by \citet{ivison02} and
\citet{richards00}, respectively. A new reduction of the HDF-N 1.4-GHz VLA data (Morrison et al.,
in preparation) has also recently featured in \citet{pope06}.

We describe our observations in \S2, the data reduction in \S3, the extraction
of sources in \S4, present differential source counts for all three fields
in \S5 and discuss our findings in \S6. Throughout we adopt a cosmology, with
$\Omega_m=0.3$, $\Omega_\Lambda=0.7$ and $H_0=70$\,km\,${\mathrm{s}^{-1}}$\,Mpc$^{-1}$.

\section{Observations}

\begin{table*}
\begin{center}
\caption{Observing dates, included VLA configurations, total integration times (target and
calibrators), 1$\sigma$ rms noise and synthesised beam parameters ({\sc fwhm}) for all fields.}
\begin{tabular}{ccccccc} \hline
Field & Observing dates & Project code & Configurations & Integration time & 1$\sigma$ rms noise  & \multicolumn{1}{c}{Synthesised beam} \\
      &                 & & & (hr)         &($\mu$Jy\,beam$^{-1}$) & (arcsec$^2$) \\ \hline
8-mJy Survey Field & 2001 Jan -- 2002 Mar & AD432, AI088, AI098 & `A',`B' & 75 & 4.6 & $1.26\times 1.32$ \\
ELAIS N2 & 2001 Jan -- May & AD432, AI091 & `A',`B' & 20 & 9.6 & $1.48\times 1.45$ \\
HDF-N & 1996 Nov -- 1996 Dec & AR368 & `A' & 45 & 5.8 & $1.52\times 1.51$ \\ \hline
\end{tabular}
\label{obstab}
\end{center}
\end{table*}

Deep, high-resolution, wide-field radio images of three northern legacy fields
were obtained using the NRAO's\footnote{NRAO is operated by Associated
Universities Inc., under a cooperative agreement with the National Science
Foundation.} VLA. These were the `8-mJy Survey field' located within the
Lockman Hole \citep{ivison02}, ELAIS N2 \citep{ivison02} and the HDF-N
\citep{richards00}.

For the 8-mJy Survey region, we used the VLA to acquire 75\,hr of data with a
4:1 ratio of time spent in the `A' and `B' configurations. For ELAIS N2, we
acquired 20\,hr of data, with a 3:1 ratio of A:B. The relatively low
integration time was due to an unresolved $\sim$120-mJy radio source, immediately
evident in the field, and sufficiently bright that dynamic-range limitations
($\ls$5,000 for the VLA) would never justify a longer integration. For the HDF-N,
we gathered all the available data from the archive, a total of 45\,hr, all in `A'
configuration\footnote{An additional 8\,hr of data was contained in the archive,
but as this was recorded several weeks before the bulk of the data and with
a longer (10\,s) integration time we did not include it in our analysis.}.
Observing parameters for each field are summarised in Table~\ref{obstab}.

Data were recorded every 5\,s (3.3\,s for the HDF-N) across two bands,
these having lower-edge frequencies of 1365 and 1435\,MHz. The correlator was configured
(mode `4') to produce seven 3.125-MHz channels per band in both right and left-circular
polarisations (RCP \& LCP). This is a commonly-used observing mode at low frequencies,
the chosen bands being relatively free of radio-frequency interference (rfi) and the use
of multiple channels per band greatly reducing the effects of bandwidth smearing.
During each day's observations, approximately 300\,s were spent on 3C\,48 (0137+331)
for ELAIS N2, or 3C\,286 (1331+305) for the Lockman Hole and the HDF-N; these data were
used to set the flux density
scale. The nearby, unresolved source, 1035+564, was used as the phase,
amplitude and bandpass calibrator in the Lockman Hole, and was observed for
100\,s every hour. Similarly, scans of 1625+415 were used to calibrate the ELAIS N2
data, and 1313+675 for the HDF-N. The remainder of the time was spent integrating on the
target fields themselves, typically for 45--50\,min between
calibration scans. When 1331+305 was not observed, the flux
density scale was set using boot-strapped flux densities for 1035+564.

\section{Data reduction}

\begin{figure}
\begin{center}
\includegraphics[scale=0.4]{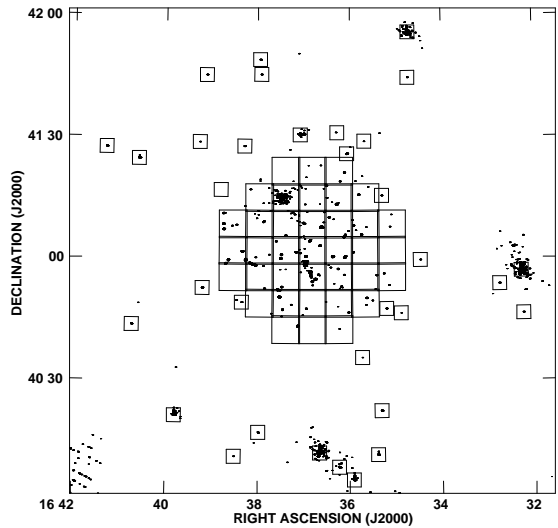}
\caption{Wide-field contour map of the ELAIS field demonstrating the location of the
central 37 facets that are used to map the VLA primary beam as well as the many other
(smaller) facets that were simultaneously mapped in order to remove the response of
sources beyond the primary beam. Contours outwith a facet are imaging artefacts and
not real sources.}
\label{facetfig}
\end{center}
\end{figure}

The reduction of the data described in \S2 followed a simple but time-intensive
formula; all data were reduced using the NRAO {\sc aips} package. First, raw data
were loaded using {\sc fillm} with {\sc douvcomp=$-$1} to minimise dynamic-range
issues, with data weights calculated based on the nominal sensitivity
and integration time. Corrupted data were identified visually and flagged using
a variety of tasks. Flux calibration proceeded in the standard fashion using
{\sc calib}, antenna gain correction factors for all sources being derived from
the flux calibrator data, 3C\,48 or 3C\,286, where appropriate. The bandpass
shape was calibrated using the unresolved calibrator, once per scan. The
solutions so found were then interpolated onto the target data, thus taking into
account any time variability of the antenna bandpasses. Finally, antenna-based
amplitude and phase corrections were found for each calibrator scan and again
interpolated onto the target data.

Due to the very large size of the datasets and the associated computing times
involved in their mapping and self-calibration, it was desirable to reduce the
amount of data as much as possible without degrading the final maps. This was
achieved using {\sc ubavg} to average data from different days together, the amount
of time averaging set, per baseline, by the requirement that the peak flux density of
sources at the edge of the field were reduced by less than 1~per~cent due to
time-averaged smearing. For those datasets where more than one VLA
configuration was present, each configuration was first averaged in this manner
before being combined together into a single file.

The calibrated data were imaged and deconvolved with {\sc imagr}. In all cases
a uniform data weighting scheme was used, with the {\sc robust} parameter set to
zero. Making a single map
of each, very large, field was impractical and so instead we followed the methodology
of \citet{owen05}, the primary beam (having an extent of $\sim$0.5$^{\circ}$ for a
25-m antenna at 1.4~GHz) being approximated with a total of 37
reasonably-sized (1024 $\times$ 1024~pixel) maps (or `facets'). Additional
maps were also made of sources detected beyond the main lobe of the primary beam,
this being necessary in order to remove their sidelobes which would otherwise
protrude into the area of the sky covered by the central facets; the {\sc clean}
components from these sources could then also be included in subsequent
self-calibration. An illustration of this technique is given in Fig.~\ref{facetfig}
for the ELAIS field. Areas of {\sc clean}ing were restricted using boxes placed
around sources, this also making it easier to use the {\sc clean} components
when self-calibrating. This was performed using {\sc calib} after each mapping
run, solving ultimately for both amplitude and phase, but initially in phase only.
The shortest solution intervals (after averaging of data in {\sc ubavg}) could be as
low as $\sim$1~min, but were generally longer for the 8-mJy Survey region (Lockman Hole)
as there are fewer bright sources in this field.

\begin{figure*}
\begin{center}
\includegraphics[scale=0.6]{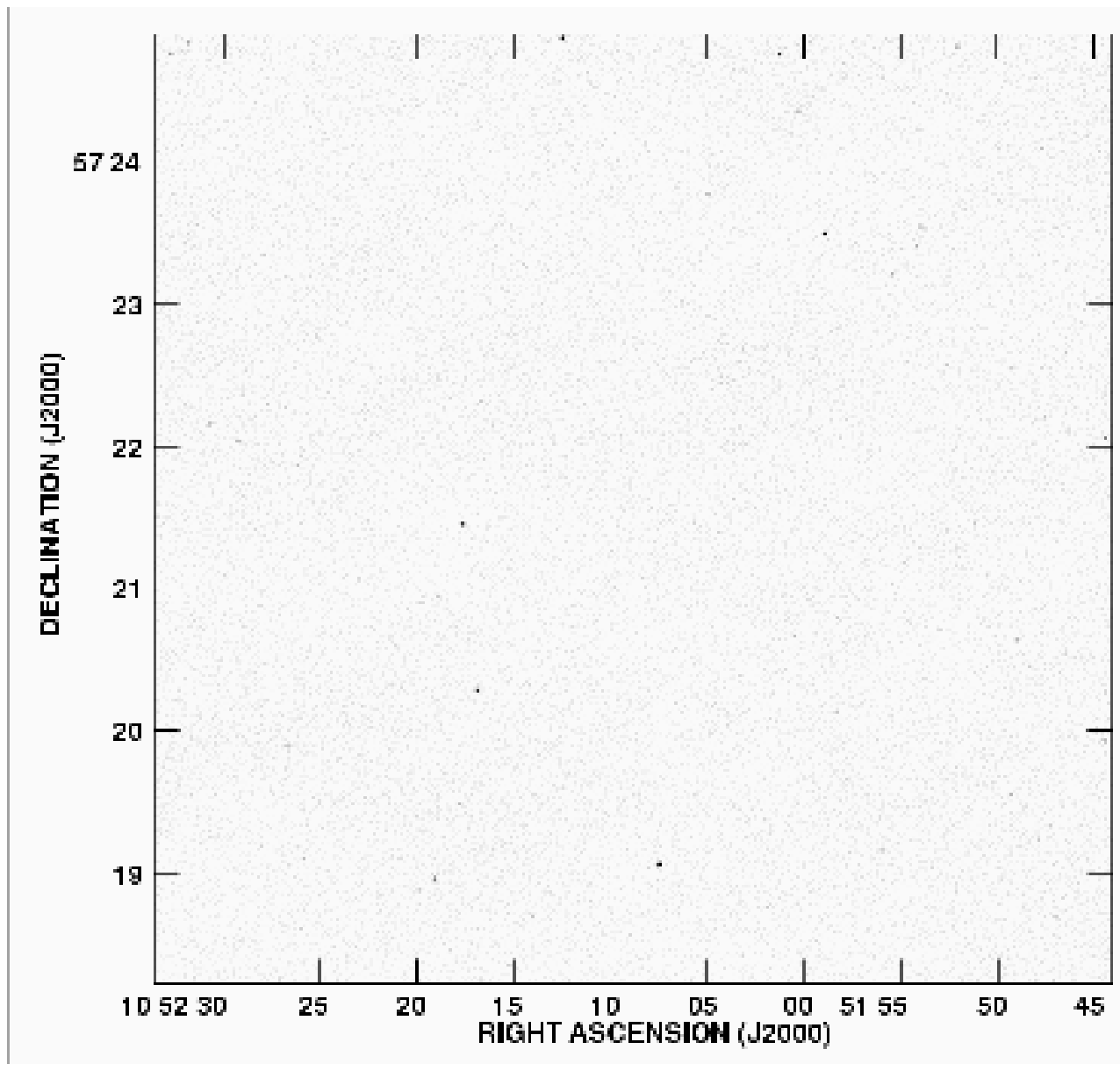}
\includegraphics[scale=0.6]{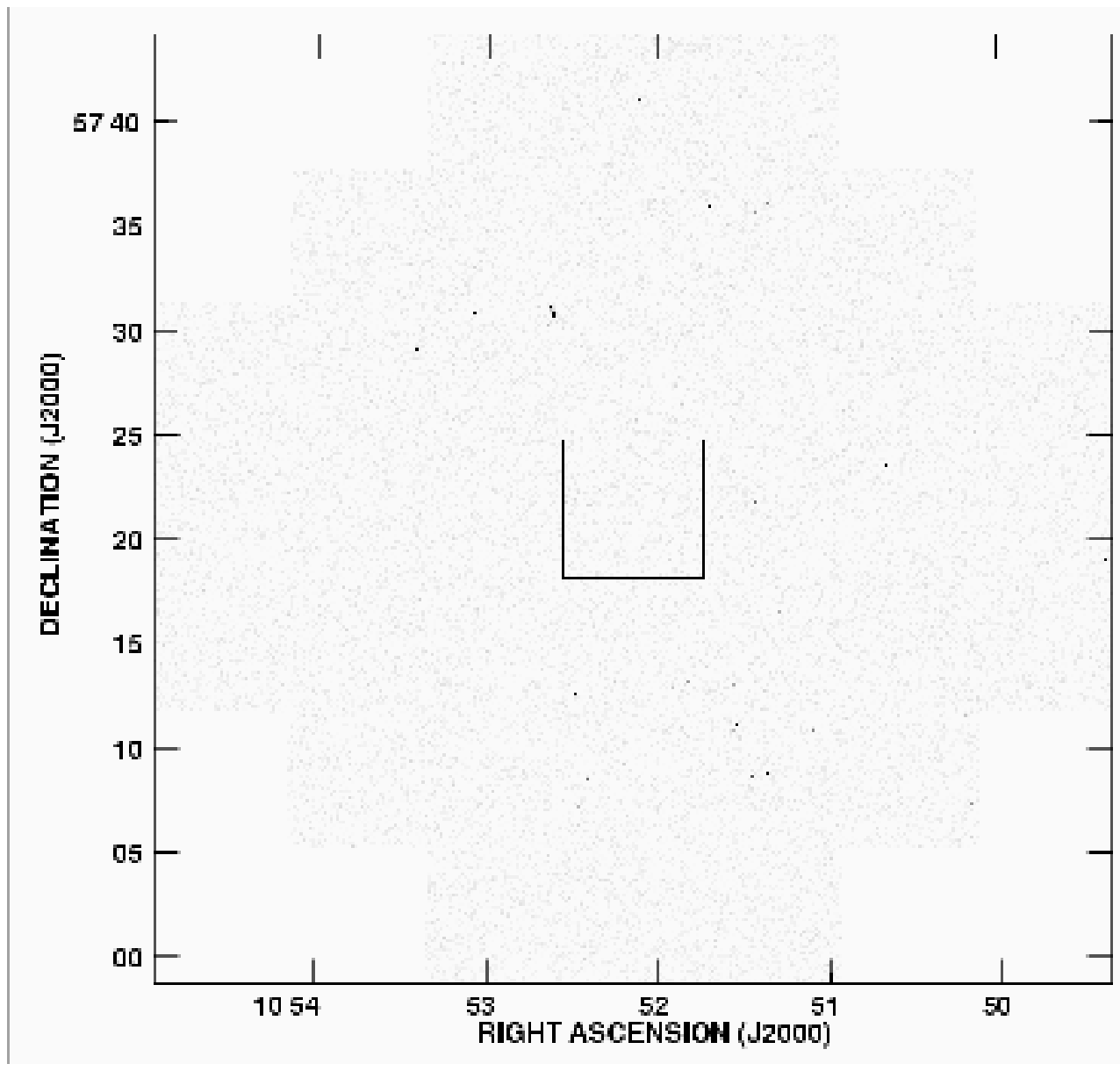}
\includegraphics[scale=0.6]{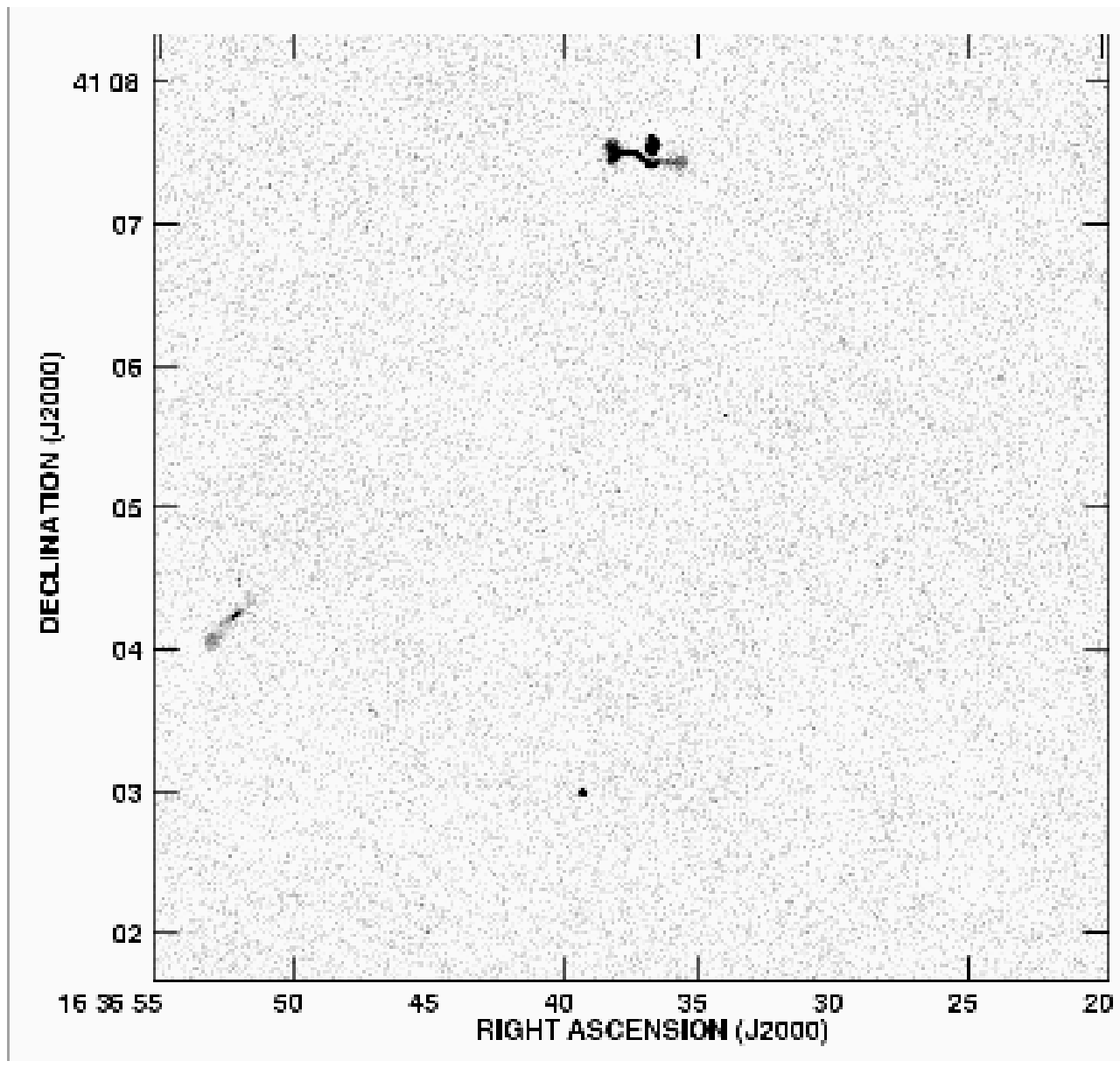}
\includegraphics[scale=0.6]{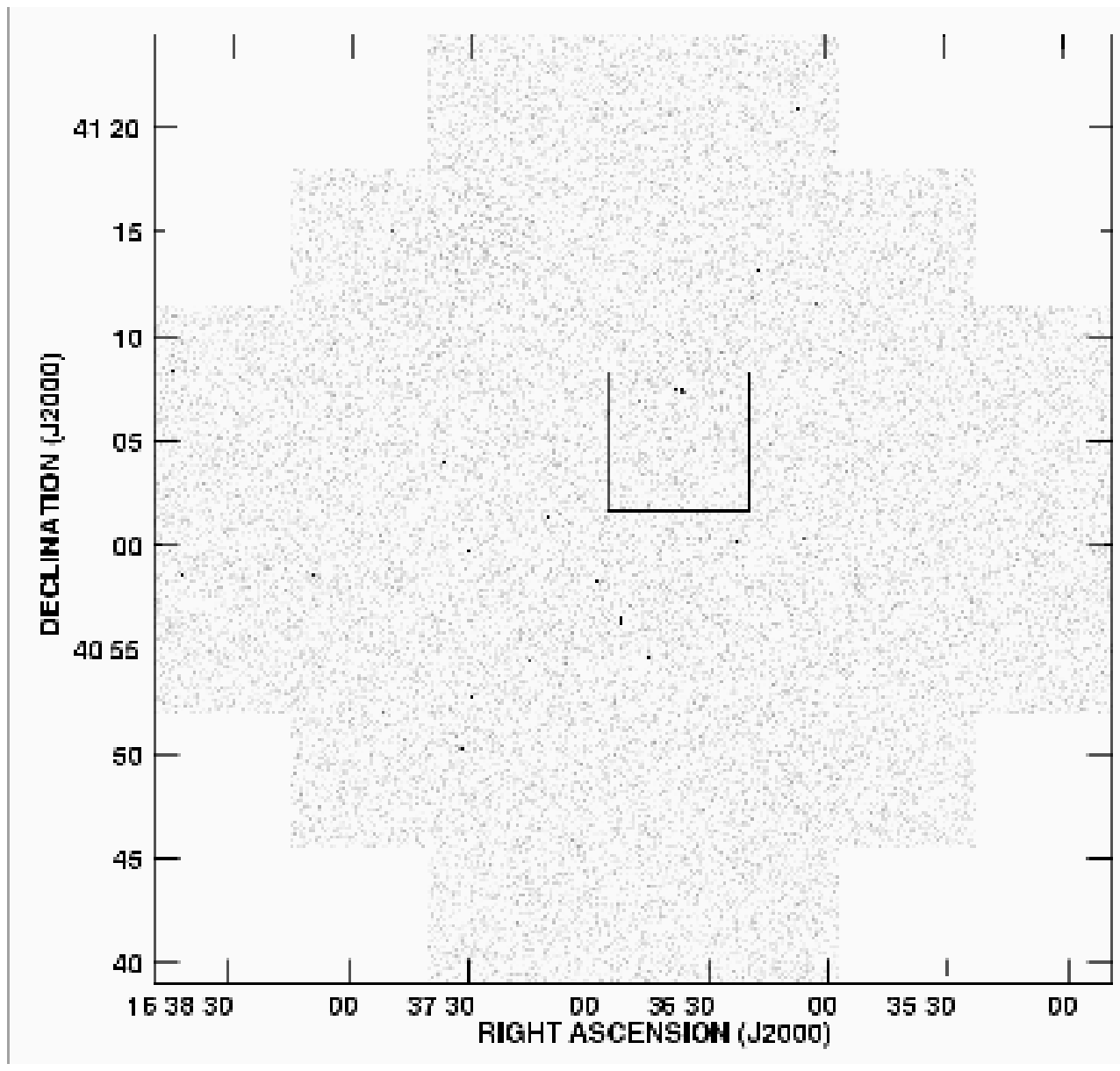}
\includegraphics[scale=0.6]{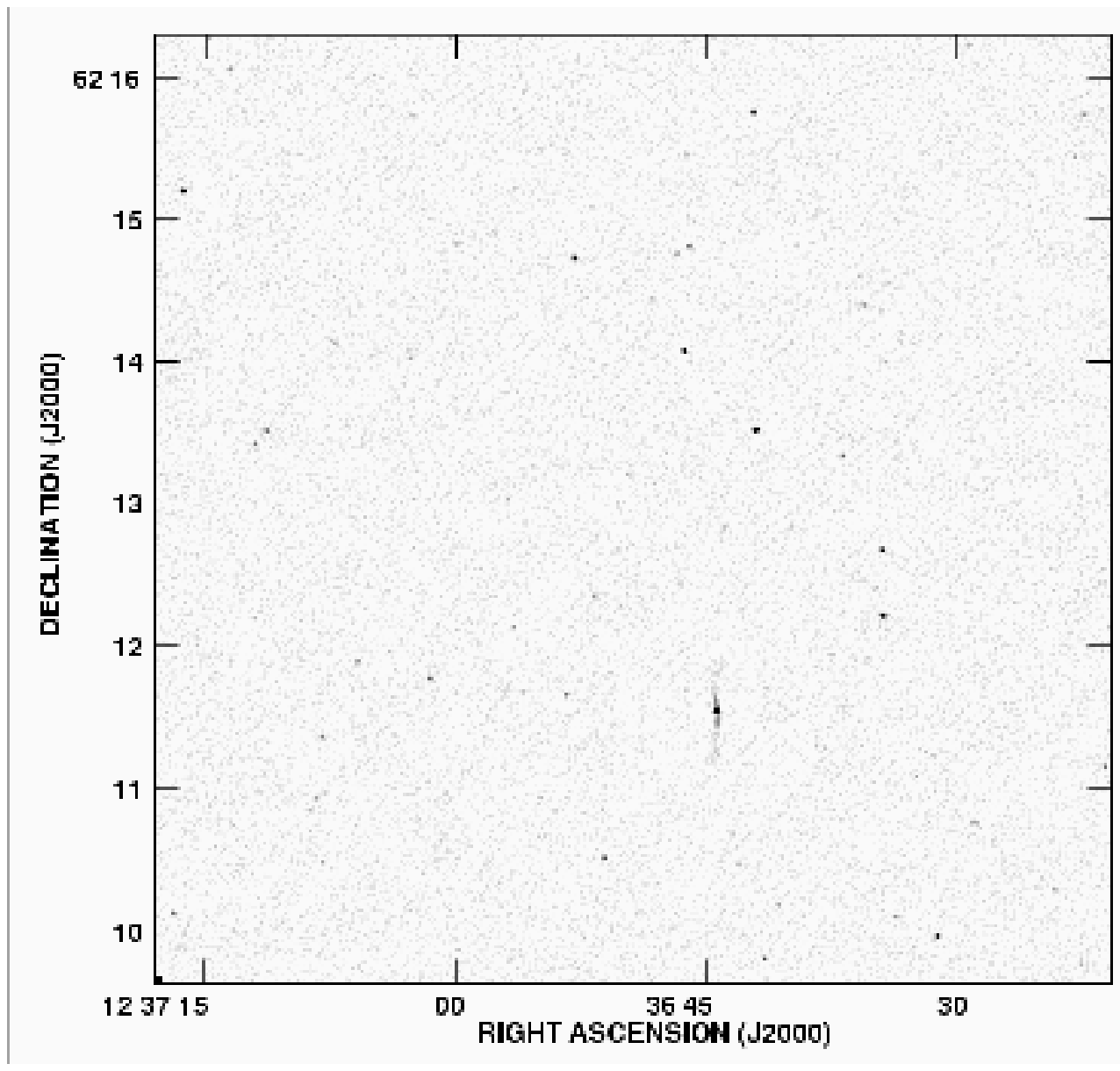}
\includegraphics[scale=0.6]{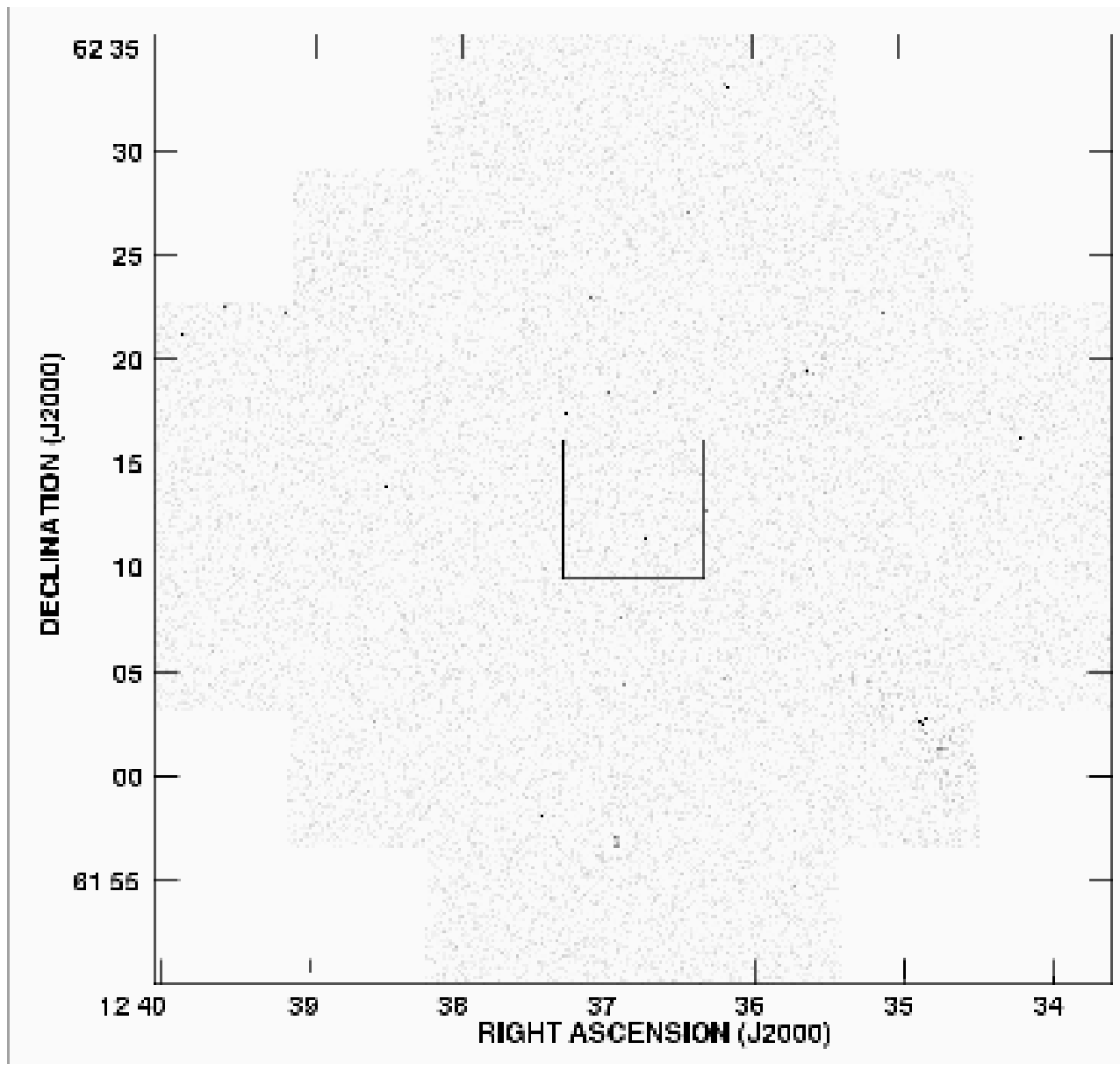}
\caption{Stokes $I$ maps of of the 8-mJy Survey field (top), ELAIS N2 field (middle) and the
Hubble Deep Field North (bottom). Right: full mosaic, with a diameter of 45\,arcmin. Left:
small $400 \times 400$-arcsec portion of the area boxed in the panels on the right.
Greyscales are plotted from 0--100\,$\mu$Jy\,beam$^{-1}$.}
\label{maps}
\end{center}
\end{figure*}

Mapping and self-calibration were performed separately on the RCP and
LCP data for all fields (excepting ELAIS N2 which is the least sensitive of our
datasets) in order to remove problems associated with the VLA `beam
squint'. At the point at which it was determined that further processing
was not necessary, the RCP and LCP maps were combined to produce Stokes $I$.
The final maps are shown in Fig.~\ref{maps}, where all 37 Stokes
$I$ facets have been combined into a single image of each field; in each
case a model of the VLA primary beam has been used to correct the
brightness of each pixel. Also shown are zooms of a $400 \times 400$-arcsec$^2$
region of each map, with this area shown on the full-field maps as a box.

We measured the 1$\sigma$ rms noise in the maps at various points in the maps
away from bright sources and obvious image artefacts caused by residual calibration
and deconvolution errors. Such problems are well illustrated by the full primary-beam
map of the HDF-N in Fig.~\ref{maps}. A number of sources can be seen that are clearly the
source of enhanced noise, particularly the two extended sources in the southern half of the
map. That near $\mathrm{12^h\,37^m, +61^{\circ}\,57^{\prime}}$ has a particularly complicated
structure and the lack of short spacings in this `A' configuration-only data almost
certainly results in larger-scale extended structure being missed, making accurate
deconvolution impossible. The second source, near $\mathrm{12^h\,37^m,
+61^{\circ}\,57^{\prime}}$ is the brightest source in the field, with a peak brightness
of 25.3\,mJy\,beam$^{-1}$, as well as having one of the most complex structures in the
field. Another extended source located beyond the primary beam towards the North-East
is responsible for the residual sidelobe structure in this quadrant of the map.
However, for all fields the measured noise was close to the theoretical value
as measured using the VLA exposure calculator\footnote{http://www.vla.nrao.edu/astro/guides/exposure/}.
The rms noise for each field was: 8-mJy Survey
(4.6\,$\mu$Jy\,beam$^{-1}$), ELAIS N2 (9.6\,$\mu$Jy\,beam$^{-1}$) and HDF-N
(5.8\,$\mu$Jy\,beam$^{-1}$). These measurements are also shown in Table~\ref{obstab}
along with the synthesised beam parameters.

\begin{figure}
\begin{center}
\includegraphics[scale=0.4]{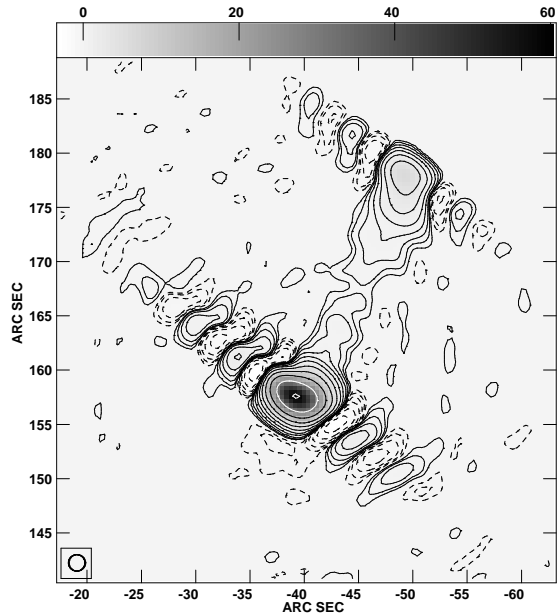}
\caption{Undeconvolvable sidelobes around the brightest source in the HDF-N; these
point directly towards the phase centre of the observations. The greyscale is given in
mJy\,beam$^{-1}$ and the synthesised beam ({\sc fwhm}) is illustrated at bottom-left. Axes
give the offset in arcsec from the centre of the facet.}
\label{sidelobe}
\end{center}
\end{figure}

A surprising feature of the maps for all three fields was the presence of sidelobes
associated with the brighter sources ($\ga$1~mJy\,beam$^{-1}$)
which proved to be undeconvolvable. These were also noted, for the HDF-N, by \citet{richards00}
who suggested that the cause was a correlator problem associated with the
3.3-s integration time. In common with these authors we find that the sidelobes
are characterised by being radially orientated towards the phase centre of the observations
and have an amplitude of $\sim$ten~per~cent. In addition, the sidelobe pattern is noticeably
stronger on the side of the source nearest the phase centre. An example of these sidelobes
is shown in Fig.~\ref{sidelobe} for the brightest source in the HDF-N. The sidelobes only
appear to affect the maps in an area immediately around each source,
the theoretical rms noise being achieved elsewhere in the maps, and therefore we
conclude that their effects on the results are not particularly pernicious.

\section{Source extraction}

\begin{table*}
\begin{center}
\caption{Extract (first ten sources) from the HDF-N catalogue, the full listing of which (along with the 8-mJy and ELAIS N2 fields)
is available online. Quantities in brackets give the 1$\sigma$ error. The third column gives the distance of the source from the
phase centre. Sources are listed in order of Right Ascension. Unresolved sources have no entries in the size columns and have integrated
flux densities equal to the peak flux density. The final column includes a note signifying whether the source is so extended that its
flux densities have been estimated using {\sc tvstat} (`E'), or whether two Gaussians have been grouped together to form a single source
(`D').}
%\begin{tabular}{llcr@{ }lr@{ }lc@{ @ }rc} \hline
\begin{tabular}{llcr@{ }lr@{ }lcccl} \hline
\multicolumn{1}{c}{Right Ascension} & \multicolumn{1}{c}{Declination} & Radial distance & \multicolumn{2}{c}{Peak flux} & \multicolumn{2}{c}{Total flux}  & \multicolumn{3}{c}{Deconvolved source size} & Note \\
\multicolumn{1}{c}{(\,$^{\mathrm{h}}\,^{\mathrm{m}}\,^{\mathrm{s}}$)} & \multicolumn{1}{c}{(\,\degr\,\arcmin\,\arcsec)} & (arcmin) & \multicolumn{2}{c}{($\mu$Jy\,beam$^{-1}$)} & \multicolumn{2}{c}{($\mu$Jy)} & Major (mas) & Minor (mas) & P.A. (\degr)  & \\ \hline
12:33:50.552 (0.027) & 62:19:35.36 (0.17) & 21.8 &   333&(52) &     434&(89) & 1.9 & 0.0 &  86.1 & \\
12:33:51.801 (0.004) & 62:16:07.57 (0.01) & 20.9 &  2206&(45) &        &     &     &     &       & \\
12:34:00.031 (0.035) & 62:12:22.08 (0.12) & 19.7 &   192&(36) &        &     &     &     &       & \\
12:34:03.510 (0.009) & 62:14:20.52 (0.03) & 19.4 &   637&(33) &        &     &     &     &       & \\
12:34:09.917 (0.033) & 62:03:57.34 (0.15) & 20.7 &   237&(45) &        &     &     &     &       & \\
12:34:10.672 (0.015) & 62:06:15.76 (0.06) & 19.7 &   441&(36) &     461&(53) & 0.7 &  0.3&  81.1 & \\
12:34:11.238         & 62:16:16.78        & 18.7 &  2349&     &    7642&     & 10.4&     &  126.5& E \\
12:34:14.475 (0.024) & 62:21:09.29 (0.13) & 19.8 &   238&(35) &     240&(50) & 0.9 &  0.0&  85.1 & \\
12:34:16.104 (0.018) & 62:07:22.18 (0.08) & 18.7 &   307&(30) &     365&(49) & 1.6 &  0.0&  46.0 & \\
12:34:16.532 (0.032) & 62:21:41.77 (0.15) & 19.8 &   195&(36) &        &     &     &     &       & \\ \hline
\end{tabular}
\label{catalogue}
\end{center}
\end{table*}

The process of extracting the position, flux density and shape/size of each source in
the data was designed to be as automated as possible.
An initial source list was first formed using the {\sc aips} task {\sc sad}.
This searches for areas in maps which lie above a certain threshold (`islands') and
fits one or more (up to a maximum of four) two-dimensional Gaussians
to these. It subsequently estimates the source size and orientation by deconvolving
the CLEAN beam from the Gaussian fit. {\sc sad} was not run on the maps produced by
{\sc imagr} but on signal-to-noise ratio (SNR) maps which were created by dividing each
{\sc imagr} map with its noise map, this having been constructed using the {\sc aips}
task {\sc rmsd}. In this manner it was possible to restrict {\sc sad} to
search for only those sources that had a peak flux density in excess of 4$\sigma$,
where $\sigma$ is the {\em local} rms noise. The noise in the maps, whilst for the
most part uniform, does vary with position, mainly due to residual calibration and
deconvolution errors as well as the afore-mentioned radial sidelobe pattern.

A small `postage-stamp' map was produced at the position of each
fitted model component, with the model Gaussian itself also plotted. By visually inspecting
each map it was possible to remove components that obviously did not correspond to a real
source, but instead to a deconvolution artefact or sidelobe. In this first stage of model
fitting, many components were also fitted to large extended sources whose size, shape,
flux and position could not be reasonably approximated by one or more Gaussians.
With both of these component types removed from further consideration in the automated
source extraction process, the remaining components were model fitted again, this time using
{\sc jmfit}.

From this point, close pairs (where the sum of the major axes was greater than or equal to the
inter-component distance) were fitted simultaneously. The area made available to
{\sc jmfit} was chosen per source, again based on the component sizes as well as the
inter-component distances where appropriate. Using the geometric mean of the major and
minor axes, those components where the undeconvolved component size was smaller than the
restoring beam (corrected for bandwidth smearing) were considered to be unresolved and their
sizes held fixed at the size of the (smeared) restoring beam. This is the case for approximately
half the sources that appear in the final catalogues, for each field.

The results of this model fitting
were then used to reject all source components with a peak flux density below 5$\sigma$. A final
round of model fitting was then performed, this time on the total intensity maps (not the SNR maps)
with corrections made both for the attenuation of the primary beam and the effect of bandwidth smearing.
At all times the model fitting was performed on the individual map facets and not on the combination
maps shown in Fig.~\ref{maps}. Due to this there are occasions when a source close to the
map edge is detected in more than one facet. If so, the fit from the map where the source lies
closest to the centre of the map is retained and the others discarded.

For those sources that could not be represented by Gaussians (18, 10 and 7 in the ELAIS, Lockman Hole
and HDFN maps respectively), their total flux densities were measured from the maps by summing all the
pixels adjudged, by eye, to make up the source (using {\sc tvstat}). The position of such a source
(marked with an `E' in the catalogue) corresponds to the peak, the catalogues also giving the largest
angular size of the source and its orientation (measured using {\sc imdist}). Of the sources fitted
with Gaussians, those components which have similar integrated flux density ratios (within a factor
of two) and which are so closely separated that the components are clearly merging are considered to
be the same source. There were also occasions when more widely separated components were clearly
associated due to a combination of their relatively small proximity given their high brightness
and/or the presence of faint extended emission between and on a line joining the two components.
In all cases these were double sources and as with the extended sources their catalogue entries
give the position of the brighter component and are marked with a `D'. The number of sources which
have been formed from multiple components is small: four in the HDFN, four in ELAIS and none in
the Lockman Hole.

Finally, a number of checks for missing sources were made, both by inspecting the residual maps
made by {\sc sad} and by making maps of each facet with the positions of the sources plotted.

The complete catalogues are available as on online supplement to this paper
and we illustrate the catalogue format for the first ten sources of the HDF-N in Table~\ref{catalogue}.
The total number of sources in each catalogue is 537 (HDFN), 200 (ELAIS) and 506 (Lockman).
We are also making our maps available to the community and these can be downloaded, in FITS format, from the
internet\footnote{ftp://ftp.roe.ac.uk/pub/adb/vla\_deep\_fields/}.

\section{Source counts}

We have used our source catalogues to derive the differential normalised source counts
for the three fields. In doing this it is crucial to correct for a number of
effects that lead to a reduction in the number of sources detected, particularly at
low flux densities. These include:
\begin{enumerate}
\item Increase in map noise due to attenuation of primary beam.
\item Reduction in source peak flux density due to bandwidth smearing.
\item Resolution bias. Sources with integrated flux densities $>5\sigma$
are missed if they are sufficiently extended that their peak flux density $<5\sigma$.
\item Reduced reliability of {\sc sad} at recovering sources when the signal to noise is low.
\end{enumerate}

In order to estimate the magnitude of these biases we have simulated a radio population
with the same characteristics as those that form our catalogues. We insert
50 sources into each of the 37 residual maps output by {\sc sad}, after these have been
corrected for the attenuation of the primary beam and the effect of bandwidth smearing.
The source integrated flux densities, $S_i$, are chosen randomly from a logarithmic distribution;
the lowest flux density is set a little below the $\sim$5$\sigma$ flux limit of the
field (20~$\mu$Jy~beam$^{-1}$ for the HDFN and Lockman Hole and 40~$\mu$Jy~beam$^{-1}$ for 
ELAIS) and the highest at a level sufficient to encompass the brightest sources used in our
source counts.

\begin{figure}
\begin{center}
\includegraphics[scale=0.34,angle=-90]{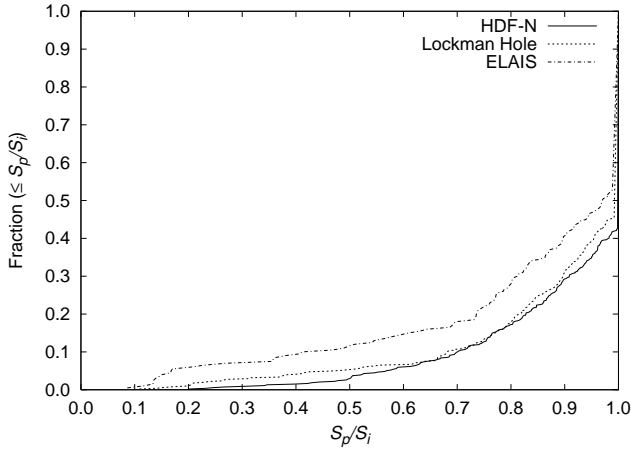}
\caption{Cumulative distribution of catalogued ratios of peak and integrated flux densities ($S_p/S_i$).
For each field it can be seen that approximately half of the sources have identically equal integrated
and peak flux densities (fitted as unresolved sources in {\sc jmfit}.)}
\label{peakdivint}
\end{center}
\end{figure}

We model our source sizes from the data themselves. The area of each modelled source is related to
the ratio of the peak flux density, $S_p$, and $S_i$. Having formed the cumulative
distribution of peak over integrated flux density for each field (Fig.~\ref{peakdivint}), we pick a
value at random and use this to calculate a modelled source size, $\theta_{mod}$, using
\begin{equation}
\theta_{mod} = \left[ \theta_{maj} \theta_{min} \left(\frac{S_i}{S_p} - 1\right) \right]^{1/2}
\end{equation}
where $\theta_{maj}$ and $\theta_{min}$ are the major and minor axes of the restoring beam. All
modelled sources are circular.

Once the flux density and size parameters are determined, sources can be added to the residual maps
(after convolution with the restoring beam) using {\sc immod}, the positions of each source also
being chosen at random. Our source extraction sequence is then run on the simulated maps, exactly
as with the observed data, and a note made as to whether a source was recovered or not.

The source counts themselves are formed by counting the number of observed sources, $N$, within bins
of integrated flux density which start at $5\sigma$ and increase by factors of 1.5. For each bin, the
smallest distance from the phase centre, $R$, necessary for all sources within that bin to be included
is noted\footnote{Due to increasingly large errors in the polynomial that describes the primary beam
correction, we do not include any sources beyond 20~arcmin in our source count calculations.}.
The source density (Jy$^{-1}$\,sr$^{-1}$) is then calculated for each bin using the area corresponding
to the radius $R$ and then Euclidean-normalised by multiplying by the average flux density of all the
sources in the bin, $\overline{S}$, to the power 2.5. The error on each bin's value is assumed to be
Poissonian i.e. $\sqrt{N}$. Only bins that contain at least nine sources are included.

\begin{figure*}
\begin{center}
\includegraphics[scale=0.6,angle=-90]{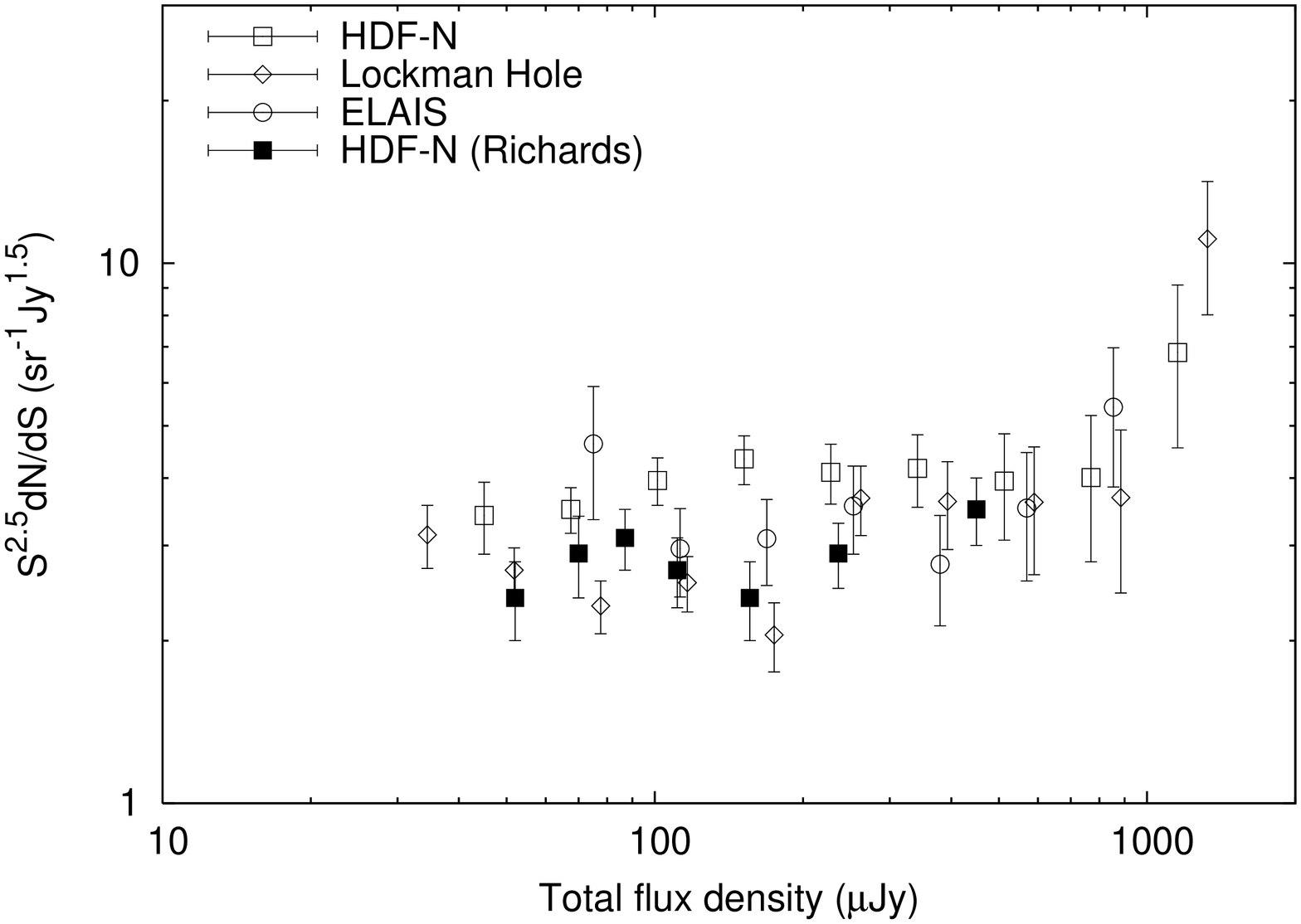}
\caption{Normalised differential source counts for the HDF-N (open squares), the Lockman Hole (open
diamonds) and ELAIS N2 (open circles). For comparison we have also plotted the \citet{richards00}
HDF-N source counts (filled squares).}
\label{scount_fig}
\end{center}
\end{figure*}

Each bin's count is then corrected for the biases described above: for all those
simulated sources which fall into the bin, the fraction that are recovered (over the same area
corresponding to the bin's radius $R$) is calculated and its inverse used as a multiplicative
corrective factor, $C$. The bin error is also scaled by this same factor. The final source counts
are given in Tables~\ref{scount_tab_hdfn}, \ref{scount_tab_lock} and \ref{scount_tab_elais} and
plotted in Fig.~\ref{scount_fig} (including the HDF-N source counts of \citeauthor{richards00}.).

%\begin{figure}
%\begin{center}
%\includegraphics[scale=0.3,angle=-90]{/harport2/adb/deep-radio/immod/hdfn/new3/immod_out.ps}
%\caption{Fraction of simulated sources recovered for the HDF-N, binned as a function of integrated flux density.
%The bins are the same as those used to form the HDF-N source counts and shown in Table~\ref{scount_tab}.}
%\label{simfrac}
%\end{center}
%\end{figure}

\begin{table*}
\begin{center}
\caption{Differential normalised source counts for the HDF-N. Tabulated are the bin upper and lower
limits ($S_l$ and $S_u$), average flux density ($\overline{S}$), number of sources per bin ($N$),
numbers of sources per bin from the \citet{richards00} catalogue ($N_{\mathrm{R00}}$), radius (in
arcmin) out to which sources were counted ($R$), bin correction factor ($C$) and the normalised
differential source counts themselves.}
\begin{tabular}{rrrrrcrr} \\ \hline
 \multicolumn{1}{c}{$S_l$} & \multicolumn{1}{c}{$S_u$} & \multicolumn{1}{c}{$\overline{S}$} & \multicolumn{1}{c}{$N$} & \multicolumn{1}{c}{$N_{\mathrm{R00}}$} & \multicolumn{1}{c}{$R$} & \multicolumn{1}{c}{$C$} & $S^{2.5}\,dN/dS$\\
($\mu$Jy) & ($\mu$Jy) & ($\mu$Jy) & & & (arcmin) & & (Jy$^{1.5}$\,sr$^{-1}$) \\ \hline
  30.0 &   45.0 &   40.2 &  43 &  1 &  9 & 2.50 & 3.41 $\pm$  0.52 \\
  45.0 &   67.5 &   56.4 & 105 & 63 & 13 & 1.41 & 3.50 $\pm$  0.34 \\
  67.5 &  101.2 &   80.9 &  99 & 96 & 16 & 1.56 & 3.96 $\pm$  0.40 \\
 101.2 &  151.9 &  122.4 &  94 & 67 & 19 & 1.35 & 4.34 $\pm$  0.45 \\
 151.9 &  227.8 &  184.5 &  63 & 59 & 20 & 1.14 & 4.10 $\pm$  0.52 \\
 227.8 &  341.7 &  266.0 &  42 & 40 & 20 & 1.04 & 4.17 $\pm$  0.64 \\
 341.7 &  512.6 &  418.8 &  20 & 22 & 20 & 1.00 & 3.95 $\pm$  0.88 \\
 512.6 &  768.9 &  629.5 &  11 & 14 & 20 & 1.00 & 4.01 $\pm$  1.21 \\
 768.9 & 1153.3 &  988.1 &   9 &  7 & 20 & 1.01 & 6.83 $\pm$  2.28 \\ \hline
\end{tabular}
\label{scount_tab_hdfn}
\end{center}
\end{table*}

\begin{table*}
\begin{center}
\caption{Differential normalised source counts for the 8-mJy Survey Region of the Lockman Hole.}
\begin{tabular}{rrrrcrr} \\ \hline
 \multicolumn{1}{c}{$S_l$} & \multicolumn{1}{c}{$S_u$} & \multicolumn{1}{c}{$\overline{S}$} & \multicolumn{1}{c}{$N$} & \multicolumn{1}{c}{$R$} & \multicolumn{1}{c}{$C$} & $S^{2.5}\,dN/dS$\\
($\mu$Jy) & ($\mu$Jy) & ($\mu$Jy) & & (arcmin) & & (Jy$^{1.5}$\,sr$^{-1}$) \\ \hline
  23.0 &  34.5 &  29.3 &  55 &  9 & 3.03 & 3.14 $\pm$  0.42 \\
  34.5 &  51.8 &  42.0 & 102 & 11 & 1.28 & 2.70 $\pm$  0.27 \\
  51.8 &  77.6 &  63.4 &  82 & 15 & 1.37 & 2.32 $\pm$  0.26 \\
  77.6 & 116.4 &  98.0 &  73 & 18 & 1.23 & 2.56 $\pm$  0.30 \\
 116.4 & 174.7 & 143.4 &  47 & 20 & 1.10 & 2.05 $\pm$  0.30 \\
 174.7 & 262.0 & 220.0 &  46 & 20 & 1.03 & 3.67 $\pm$  0.54 \\
 262.0 & 393.0 & 309.4 &  29 & 20 & 1.03 & 3.62 $\pm$  0.67 \\
 393.0 & 589.5 & 490.6 &  14 & 20 & 1.01 & 3.61 $\pm$  0.96 \\
 589.5 & 884.2 & 696.6 &   9 & 20 & 1.00 & 3.68 $\pm$  1.23 \\
 884.2 &1326.3 &1099.5 &  13 & 20 & 1.00 &11.09 $\pm$  3.07 \\ \hline
\end{tabular}
\label{scount_tab_lock}
\end{center}
\end{table*}

\begin{table*}
\begin{center}
\caption{Differential normalised source counts for ELAIS\,N2.}
\begin{tabular}{rrrrcrr} \\ \hline
 \multicolumn{1}{c}{$S_l$} & \multicolumn{1}{c}{$S_u$} & \multicolumn{1}{c}{$\overline{S}$} & \multicolumn{1}{c}{$N$} & \multicolumn{1}{c}{$R$} & \multicolumn{1}{c}{$C$} & $S^{2.5}\,dN/dS$\\
($\mu$Jy) & ($\mu$Jy) & ($\mu$Jy) & & (arcmin) & & (Jy$^{1.5}$\,sr$^{-1}$) \\ \hline
  50.0 &  75.0 &  68.1 &  13 &  7 & 3.03 & 4.63 $\pm$  1.28 \\
  75.0 & 112.5 &  95.0 &  29 & 12 & 1.67 & 2.96 $\pm$  0.55 \\
 112.5 & 168.8 & 140.2 &  30 & 14 & 1.30 & 3.09 $\pm$  0.56 \\
 168.8 & 253.1 & 209.0 &  29 & 18 & 1.41 & 3.55 $\pm$  0.66 \\
 253.1 & 379.7 & 297.2 &  19 & 19 & 1.16 & 2.77 $\pm$  0.64 \\
 379.7 & 569.5 & 450.6 &  14 & 19 & 1.06 & 3.52 $\pm$  0.94 \\
 569.5 & 854.3 & 702.9 &  12 & 20 & 1.04 & 5.41 $\pm$  1.56 \\ \hline
\end{tabular}
\label{scount_tab_elais}
\end{center}
\end{table*}

\section{Discussion}

Fig.~\ref{scount_fig} only significantly samples the so-called `upturn' in the normalised differential
source counts, the point at flux densities below approximately 1~mJy where the steep decline seen at
higher flux densities, just visible in our plot, is arrested. We find that there is general agreement
between the source counts for each of the three fields and that these are consistent with the results
found by other authors for other fields; the HDF-N source counts, for example, are extremely similar
to those found by \citet{seymour04} for the 13$^\mathrm{h}$\, {\em XMM-Newton/ROSAT} field. The counts
for the Lockman Hole are lower, especially for $S_i \la 200 \mu$Jy, but this is not altogether surprising
given that less sources were detected here than in the HDF-N, despite the Lockman Hole data being more
sensitive.

The correction factors quoted in Table~\ref{scount_tab_hdfn} are higher than
those quoted by some other authors, but it is important to compare like with like. For instance, the
correction factors of \citet{richards00} do not include the attenuation of the primary beam or
bandwidth smearing, this instead being factored into their quoted value of $dN/dS$. \citet{seymour04}, do
include primary beam and chromatic aberration effects in their correction factor
and for their lowest bin this is equal to 2.56, similar to our corresponding value for the HDF-N.
In general, source counts derived from single pointings of a radio
telescope \citep[e.g.][]{richards00,seymour04} will require large corrections to the lowest bins because
the noise is so non-uniform over the beam area. Observations which mosaic several pointings together
\citep[e.g.][]{hopkins98,bondi03} have more uniform noise and consequently lower corrections.

Most noteworthy in Fig.~\ref{scount_fig} is the striking inconsistency between the HDF-N source counts
presented in this paper and by \citet{richards00}, the new counts being significantly higher.

\subsection{Disparity with the \citet{richards00} results}

The finding that our source counts for the HDF-N are higher than those published by \citet{richards00}
is particularly interesting as the previous result has been noted for many years (beginning with
\citeauthor{richards00}) as being anomalously low; the new counts on the other hand are extremely
consistent with the results from other fields. This in itself would suggest that the source counts
presented here are more accurate, but, given the disparity, it is important to perform a careful
comparison with the original.

In terms of number of sources, our catalogue contains many more, 537 versus 371. This is partly due
to the new catalogue covering a larger area; out to the same radius (20~arcmin), \citeauthor{richards00}
finds 370 sources\footnote{The first source in the \citeauthor{richards00} catalogue is actually
28.3~arcmin from the phase centre.} compared to our 506. The remaining surplus is due to our maps
being significantly deeper and shouldn't contribute to the discrepancy in the source counts.

We note that there are 34 sources in the \citeauthor{richards00} catalogue that do not appear in
ours. We have scrutinised our maps at the positions of these sources and in the majority of cases
conclude that weak emission is present, but below our 5$\sigma$ peak cutoff. Most of these sources
have a quoted SNR of between five and seven although one, with a total flux density of 1.9~mJy
(SNR $>$ 40) is certainly not real, there being blank sky at the quoted position both in our map
and in FIRST. Different data weighting is one possible explanation for why the other sources do
not make it into our catalogue.

By identically binning the sources from both catalogues we can directly compare the source counts
from each. We have binned the \citeauthor{richards00} sources in the same manner as our own and
find, not surprisingly given the better sensitivity of the new maps, that in the lowest bins we
detect many more sources (Table~\ref{scount_tab_hdfn}). In the subsequent bins there is a general
trend for our catalogue to contribute more sources, but in two cases the \citeauthor{richards00}
catalogue has more. In general, the difference in the numbers detected is not sufficient to account
for the systematic offset. For example, at $\sim$180\,$\mu$Jy, both catalogues have similar numbers
of sources, but there is almost a factor of two difference between the source counts. We have also
calculated the integrated flux density ratios for the 334 sources common to both catalogues and find
(after removal of three outliers from \citeauthor{richards00} whose flux densities are in error
by factors $\ge$10) a mean of 1.02 and an rms of $\sim$20\%.

Ultimately, we are unable to identify the cause of the lower HDF-N source counts. However,
we note that it does not seem possible to arrive at the values of $n/n_o$ from
the information given in Table~3
of \citeauthor{richards00}. These should be obtainable by multiplying $dN/dS$ by the correction factor
and then by the average flux density to the power 2.5. If we assume that column~2 (labelled
$\langle S_{1.4}\rangle$) is the appropriate measure of flux density, we find that the numbers
obtained in this manner do not agree with and are in fact higher than those reported in Table~3
(apart from the highest flux density bin) and therefore in better agreement with our counts.

\subsection{New submm counterparts in the HDF-N}

\begin{table*}
\begin{center}
\caption{Submm sources with new radio counterparts identified in the new reduction of the HDF-N data.
Submm names and positions are taken from \citet{pope05}.}
\begin{tabular}{cccc} \\ \hline
Submm name & Submm position & Radio position & Radio flux density ($\mu$Jy) \\ \hline
GN03 (SMMJ123608$+$621147) & 12$^\mathrm{h}$36$^\mathrm{m}$08\fs9, $+$62\degr12\arcmin53\arcsec & 
12$^\mathrm{h}$36$^\mathrm{m}$08\fs665, $+$62\degr12\arcmin51\farcs26 & 34$\pm$7 \\
GN20 (SMMJ123711$+$622212) & 12$^\mathrm{h}$37$^\mathrm{m}$11\fs7, $+$62\degr22\arcmin12\arcsec &
12$^\mathrm{h}$37$^\mathrm{m}$11\fs875, $+$62\degr22\arcmin11\farcs54 & 58$\pm$16 \\ \hline
\end{tabular}
\label{submm_tab}
\end{center}
\end{table*}

\begin{figure*}
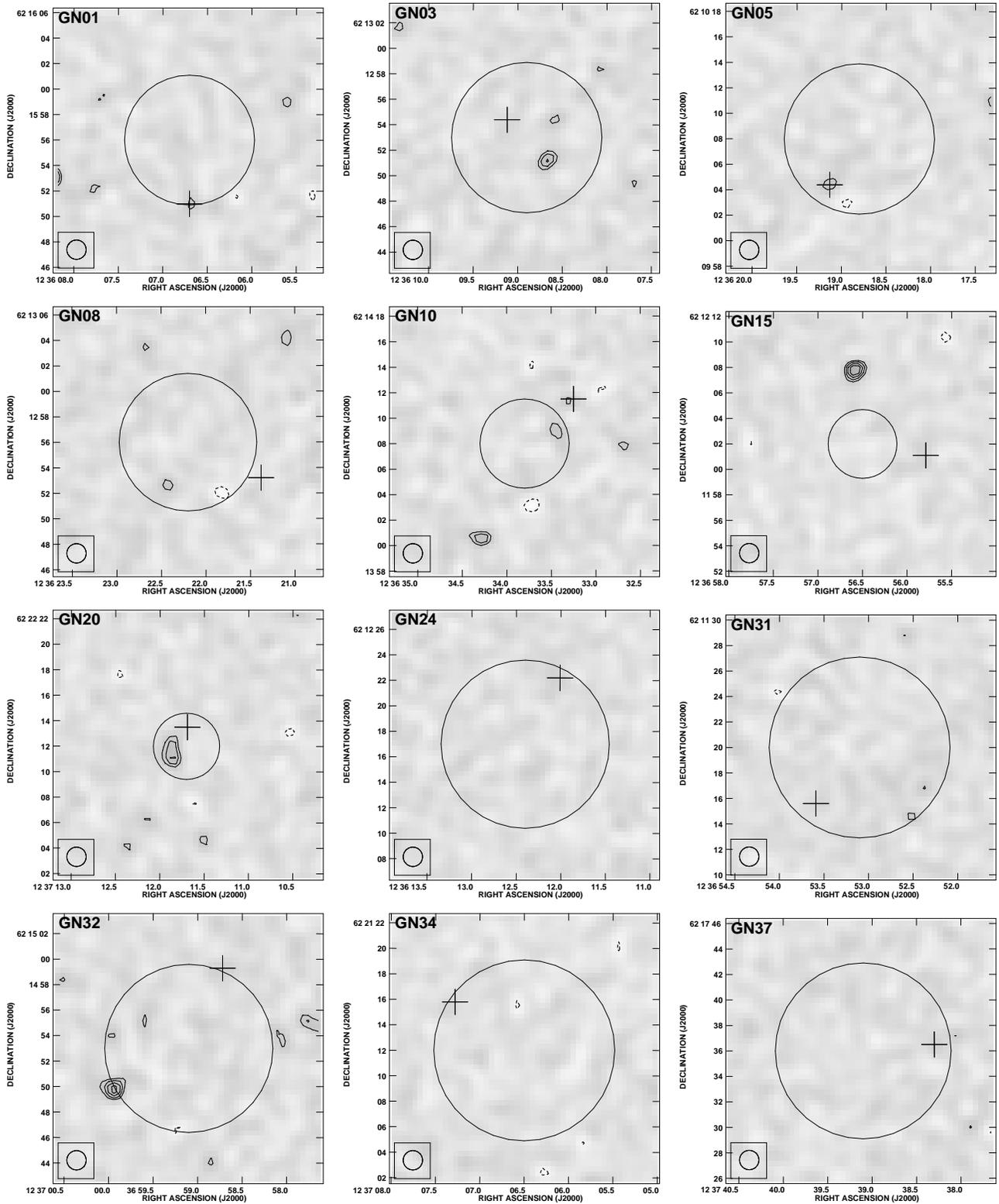

\begin{center}
\includegraphics[scale=0.3]{fig6a.ps}
\includegraphics[scale=0.3]{fig6b.ps}
\includegraphics[scale=0.3]{fig6c.ps}
\includegraphics[scale=0.3]{fig6d.ps}
\includegraphics[scale=0.3]{fig6e.ps}
\includegraphics[scale=0.3]{fig6f.ps}
\includegraphics[scale=0.3]{fig6g.ps}
\includegraphics[scale=0.3]{fig6h.ps}
\includegraphics[scale=0.3]{fig6i.ps}
\includegraphics[scale=0.3]{fig6j.ps}
\includegraphics[scale=0.3]{fig6k.ps}
\includegraphics[scale=0.3]{fig6l.ps}
\caption{VLA 1.4-GHz maps centred on the positions of the 12 submm sources in the HDF-N
described by \citet{pope05} as having an optical/near-IR, but no radio, counterpart. Each map is $20 \times
20$~arcsec$^2$ with contours plotted at $-3$, 3, 4, 5, 6 times the 1$\sigma$ rms noise. The
circle describes the 2$\sigma$ submm position error based on the signal to noise ratio
(Ivison et al., in preparation). The cross marks the position of the optical/near-IR
counterpart \citep{pope05}. Ellipses in the bottom-left corners represent the telescope beam ({\sc fwhm}).}
\label{submm}
\end{center}
\end{figure*}

An invaluable application of deep radio catalogues to cosmology has been their usefulness in
identifying the optical counterparts to submm sources detected by e.g. the Submillimetre
Common User Bolometer Array \citep[SCUBA,][]{holland99}. The much smaller beam of the
VLA at 1.4~GHz (1--2~arcsec) compared to that of SCUBA ($\sim$15~arcsec at 850~$\mu$m) and
the fact that submm sources often have associated radio emission makes it possible to get
much more accurate positions for SCUBA sources which can then be used to identify the
galaxy in the optical and infrared \citep[e.g.][]{ivison02}.

Interest in the HDF-N remains high, especially with the advent of GOODS
\citep{dickinson03}, a deep multi-wavelength survey which, amongst other things,
seeks to constrain theories of galaxy evolution. Its northern field, GOODS-N, coincides
with the HDF-N. \citet{pope05} present a recent SCUBA 850-$\mu$m catalogue of the GOODS-N
field which contains 40 sources, several of which are previously unpublished. Of these, 27
have an optical counterpart and twelve of these do not have known radio detections. In
Fig.~\ref{submm}, a $20 \times 20$~arcsec$^2$ VLA map centred on the submm position of each of
these twelve sources is plotted. A circle indicates the 2$\sigma$ positional error (SNR
dependent, Ivison et al., in preparation) and a cross the position of the optical/near-IR
identification.

Two of the submm sources (GN03 and GN20) have a clear 5$\sigma$ radio detection within
the error circle and both of these are included in our HDF-N catalogue. Of particular
interest is GN20, this previously being the brightest (21.3~mJy) submm
source without an identified counterpart in a different waveband. It is, however, clearly
detected in our map, is slightly resolved and lies only 1.3~arcsec away from the submm position.
In the case of GN03, the radio/submm positional offset is 2.4~arcsec. Positions and flux
densities for these two sources are given in Table~\ref{submm_tab}. Similar flux densities
for these sources are found by \citet{pope06}. A third source, GN32, also
has a radio source within the 2$\sigma$ error circle, but this is already present in the
\citet{richards00} catalogue.

\section{Conclusions}

In this paper we have presented radio maps and catalogues from three legacy fields, the HDF-N, the
8-mJy Survey Region of the Lockman Hole and ELAIS~N2. These are areas of the sky which
continue to be observed at all wavelengths and with a variety of telescopes, resolutions, etc;
we hope that by making our radio maps available, the continued exploitation of these fields
will be made much easier given that the protracted tasks of assembly, calibration and mapping
of the data will be not be necessary.

The data presented here have been discussed before by other authors, but the new reduction of
the HDF-N data is a marked improvement on that already published \citep{richards00} and has yielded
a new source catalogue with more and fainter sources. This has been used to find radio counterparts
to two submm sources which previously had none; both of these identifications have recently also been
made by \citet{pope06} from a separate re-analysis of the HDF-N data (Morrison et al., in preparation).
By going deeper still it may be possible to identify the radio counterparts to those submm sources in
the HDF-N which remain undetected, sources which otherwise might be considered spurious
\citep[e.g.][]{greve04}. Deeper, more reliable radio maps of the HDF-N should be available in the
near future due to further VLA observations at 1.4~GHz in the `A, `B', `C' and `D' configurations
(P.I. Glenn Morrison).

Source counts have been presented for each of our fields, down to a flux density limit of
5$\sigma$. These are broadly consistent with one another, but the observed differences (such as
the low counts seen in the Lockman Hole) might be a consequence of cosmic variance. Our most
important conclusion is that we find no evidence for the under-density of sources in the
HDF-N reported by \citet{richards00}, instead obtaining results which are entirely consistent with
those found for other fields. It has long been noted that the HDF-N source counts were
anomalously low and this finding appears to have been spurious. The recent resurgence
in interest in the HDF-N radio data, both in the re-analyses of the archived data by ourselves and
Morrison et al. (in preparation) and the on-going effort to add new data, should confirm our
results and allow the source counts to be extended to still fainter flux densities.

\section*{Acknowledgments}

ADB would like to thank Nuria Lorente for her help in getting
various perl scripts and IDL programs working and Nick Seymour
for his thoughts on source counts.
RJI acknowledges kind and patient help over the years from Ernie
Seaquist, Frazer Owen, Bob Becker, Chris Carilli and Rick White.
We thank the anonymous referee for many suggestions which greatly
improved this manuscript.

\bibliographystyle{mnras}
\bibliography{deep}

\begin{thebibliography}{}

\bibitem[\protect\citeauthoryear{{Bondi} et~al.}{{Bondi}
  et~al.}{2003}]{bondi03}
{Bondi} M. et~al., 2003, \aap, 403, 857

\bibitem[\protect\citeauthoryear{{Condon} et~al.}{{Condon}
  et~al.}{1998}]{condon98}
{Condon} J.~J., {Cotton} W.~D., {Greisen} E.~W., {Yin} Q.~F., {Perley} R.~A.,
  {Taylor} G.~B.,  {Broderick} J.~J., 1998, \aj, 115, 1693

\bibitem[\protect\citeauthoryear{{Condon} et~al.}{{Condon}
  et~al.}{2003}]{condon03}
{Condon} J.~J., {Cotton} W.~D., {Yin} Q.~F., {Shupe} D.~L., {Storrie-Lombardi}
  L.~J., {Helou} G., {Soifer} B.~T.,  {Werner} M.~W., 2003, \aj, 125, 2411

\bibitem[\protect\citeauthoryear{{Dickinson}, {Giavalisco} \& {The Goods
  Team}}{{Dickinson} et~al.}{2003}]{dickinson03}
{Dickinson} M., {Giavalisco} M.,  {The Goods Team} , 2003, in The Mass of
  Galaxies at Low and High Redshift, p. 324

\bibitem[\protect\citeauthoryear{{Donley} et~al.}{{Donley}
  et~al.}{2005}]{donley05}
{Donley} J.~L. et~al., 2005, \aj, 129, 220

\bibitem[\protect\citeauthoryear{{Dunlop}}{{Dunlop}}{1998}]{dunlop98}
{Dunlop} J.~S., 1998, in ASSL Vol. 226: Observational Cosmology with the New
  Radio Surveys, p. 157

\bibitem[\protect\citeauthoryear{{Dunlop} \& {Peacock}}{{Dunlop} \&
  {Peacock}}{1990}]{dunlop90}
{Dunlop} J.~S.,  {Peacock} J.~A., 1990, \mnras, 247, 19

\bibitem[\protect\citeauthoryear{{Greve} et~al.}{{Greve}
  et~al.}{2004}]{greve04}
{Greve} T.~R., {Ivison} R.~J., {Bertoldi} F., {Stevens} J.~A., {Dunlop} J.~S.,
  {Lutz} D.,  {Carilli} C.~L., 2004, \mnras, 354, 779

\bibitem[\protect\citeauthoryear{{Holland} et~al.}{{Holland}
  et~al.}{1999}]{holland99}
{Holland} W.~S. et~al., 1999, \mnras, 303, 659

\bibitem[\protect\citeauthoryear{{Hopkins} et~al.}{{Hopkins}
  et~al.}{2003}]{hopkins03}
{Hopkins} A.~M., {Afonso} J., {Chan} B., {Cram} L.~E., {Georgakakis} A.,
  {Mobasher} B., 2003, \aj, 125, 465

\bibitem[\protect\citeauthoryear{{Hopkins} et~al.}{{Hopkins}
  et~al.}{1998}]{hopkins98}
{Hopkins} A.~M., {Mobasher} B., {Cram} L.,  {Rowan-Robinson} M., 1998, \mnras,
  296, 839

\bibitem[\protect\citeauthoryear{{Huynh} et~al.}{{Huynh}
  et~al.}{2005}]{huynh05}
{Huynh} M.~T., {Jackson} C.~A., {Norris} R.~P.,  {Prandoni} I., 2005, \aj, 130,
  1373

\bibitem[\protect\citeauthoryear{{Ivison} et~al.}{{Ivison}
  et~al.}{2002}]{ivison02}
{Ivison} R.~J. et~al., 2002, \mnras, 337, 1

\bibitem[\protect\citeauthoryear{{Lockman}, {Jahoda} \& {McCammon}}{{Lockman}
  et~al.}{1986}]{lockman86}
{Lockman} F.~J., {Jahoda} K.,  {McCammon} D., 1986, \apj, 302, 432

\bibitem[\protect\citeauthoryear{{Morganti} et~al.}{{Morganti}
  et~al.}{2004}]{morganti04}
{Morganti} R., {Garrett} M.~A., {Chapman} S., {Baan} W., {Helou} G.,  {Soifer}
  T., 2004, \aap, 424, 371

\bibitem[\protect\citeauthoryear{{Muxlow} et~al.}{{Muxlow}
  et~al.}{2005}]{muxlow05}
{Muxlow} T.~W.~B. et~al., 2005, \mnras, 358, 1159

\bibitem[\protect\citeauthoryear{{Owen} et~al.}{{Owen} et~al.}{2005}]{owen05}
{Owen} F.~N., {Keel} W.~C., {Ledlow} M.~J., {Morrison} G.~E.,  {Windhorst}
  R.~A., 2005, \aj, 129, 26

\bibitem[\protect\citeauthoryear{{Pope} et~al.}{{Pope} et~al.}{2005}]{pope05}
{Pope} A., {Borys} C., {Scott} D., {Conselice} C., {Dickinson} M.,  {Mobasher}
  B., 2005, \mnras, 358, 149

\bibitem[\protect\citeauthoryear{{Pope} et~al.}{{Pope} et~al.}{2006}]{pope06}
{Pope} A. et~al., 2006, \mnras, accepted (astro-ph/0605573)

\bibitem[\protect\citeauthoryear{{Richards}}{{Richards}}{2000}]{richards00}
{Richards} E.~A., 2000, \apj, 533, 611

\bibitem[\protect\citeauthoryear{{Seymour}, {McHardy} \& {Gunn}}{{Seymour}
  et~al.}{2004}]{seymour04}
{Seymour} N., {McHardy} I.~M.,  {Gunn} K.~F., 2004, \mnras, 352, 131

\bibitem[\protect\citeauthoryear{{White} et~al.}{{White}
  et~al.}{1997}]{white97}
{White} R.~L., {Becker} R.~H., {Helfand} D.~J.,  {Gregg} M.~D., 1997, \apj,
  475, 479

\end{thebibliography}

\end{document}